\documentstyle{llncs}
\newcommand{\ba}{\[\begin{array}{r@{\qquad}rcl}}
\newcommand{\ea}{\end{array}\]}

\newenvironment{textcode}{\noindent\hrulefill}{\hrulefill\bigskip}

\newcommand{\splname}{\textsc{SL}}
\newcommand{\splsrchost}{\textsc{Caml}}
\newcommand{\spltarget}{\textsc{Caml}}

\newcommand{\eg}{\textit{e.g.}}

\newcommand{\aka}{\textit{a.k.a.}}

\newcommand{\concept}[1]{{\emph{#1}}}

\newcommand{\ttbox}[1]{\mbox{\tt{#1}}}
\newcommand{\bfbox}[1]{\mbox{\bf{#1}}}

\newcommand{\alt}{\,\,\vert\,\,}
\newcommand{\arrow}{\rightarrow}

\newcommand{\circarrow}{\circ\!\!\!\arrow}

\newcommand{\infer}[2]{\displaystyle{\frac{#1}{#2}}}

\newcommand{\tjudge}[3]{{#1} ~\vdash~ {#2} \mbox{:} {#3}}

\newcommand{\dbllbracket}{[\![}
\newcommand{\dblrbracket}{]\!]}
\newcommand{\transbracket}[1]{\dbllbracket ~{#1}~ \dblrbracket}

\newcommand{\comment}[1]{}

\newcommand{\chcstatessep}{\bfbox{\tt |}}
\newcommand{\chcstates}[2]{{#1} \bfbox{\tt |} \cdots \chcstatessep {#2}}
\newcommand{\letstate}[2]{\ttbox{let}~{#1}~\ttbox{in}~{#2}}
\newcommand{\transstate}[2]{\transbracket{#1} ~~{#2}~}

\begin{document}



\title{From Syntactic Theories to Interpreters: \\
       A Specification Language and Its Compilation}
\author{Yong Xiao\inst{1}
       \hspace{1cm} Zena M. Ariola\inst{1} \hspace{1cm} Michel Mauny\inst{2}}

\institute{University of Oregon, Eugene, OR 97403\thanks{Supported by
NSF grant CCR-9624711} 
\and
INRIA-Rocquencourt and New York University}
\maketitle

\begin{abstract}
Recent years have seen an increasing need of high-level specification
languages and tools generating code from specifications. 
In this paper, we
introduce a specification language, {\splname}, which is tailored to
the writing of syntactic theories of language semantics. More specifically,
the language supports specifying primitive notions such as dynamic
constraints, contexts, axioms, and
inference rules. 
We also introduce
a system which generates
interpreters from {\splname} specifications.
A prototype system is implemented and has been tested on
a number of examples, including a syntactic theory for Verilog. 

\end{abstract}

\section{Introduction}
\label{secintro}

%
Syntactic theories have been developed to reason about many aspects of
modern programming languages\cite{cyclic,cbneed,monadicState,
monadic-ml,verilog}. Having roots in the $\lambda$-calculus,
these theories rely on transforming source programs to other source programs.
Only the \emph{syntax} of the programming language is relevant.


Experience
shows that the development of such theories is 
error-prone. 
In order to ensure that the theories are sensible, 
many properties need to be checked. For example,
we need to know if
 we have enough rules to rewrite a program to
its value, if the type of a program is preserved during evaluation,
and  whether each program has a unique value.
In many situations, the proofs of these properties do
not require deep insight. In fact, 
many purported proofs suffer from being incomplete, and 
usually the missed case is the
problematic one. Thus, we think that in order to rely on syntactic
theories, it is of mandatory importance to design 
tools that support their development. The work described in this paper
offers a first step towards that direction. 

We introduce the specification language {\splname}, 
which can directly reflect the primitive notions of syntactic theories
such as 
evaluation contexts and dynamic constraints. 
An experimental system has been implemented that generates interpreters
from {\splname} specifications. Currently the generated interpreters
are programs in {\spltarget}\cite{Camlsystem} which is a dialect in the ML family. 
Various examples
have been tested, including the operational semantics of core-ML 
(lambda-calculus with built-in  operations, store operations, 
and exception handling), type inference for core-ML, 
a syntactic theory for a store encapsulation
language\cite{monadic-ml}, and a syntactic theory for  
Verilog\cite{verilog}.



The paper
is organized as follows: 
Section~\ref{sec_overview} 
gives an overview of the {\splname} system using 
the call-by-value $\lambda$-calculus as an example. 
Section~\ref{sec_compile} 
describes some issues in compiling {\splname} programs,
such as type checking for contexts, pattern-matching,
and code generation. 
Section~\ref{sec_discussion} presents related work and concludes the
paper.

\section{Overview of the {\splname} System}
\label{sec_overview}

We introduce the call-by-value $\lambda$-calculus\cite{Plo} and show
how it is specified in the {\splname} language. 
By convention, we call the {\splname}
language the \concept{meta-language}, and call the call-by-value
$\lambda$-calculus the \concept{object-language}.

\subsection{A Syntactic Theory for Call-by-value $\lambda$-calculus}

The set of terms of the call-by-value lambda calculus is generated
inductively over an infinite set of variables (ranged over by
$x$, $y$, etc): it includes
$\lambda$-abstractions and applications:

\ba
\textit{Terms} &
        M &::=& x \alt \lambda x.M \alt MM \\
\ea

The semantics is based on the $\beta_v$ reduction rule which requires
a syntactic definition of the notion of value:

\[
\begin{array}{l@{\quad}rll}
\textit{Values} & V &::=& \lambda x.M \\
\beta_v & (\lambda x.M) V & \rightarrow&  M[x:=V] 
\end{array}
\]
\noindent
where $M[x:=V]$ is the term resulting from substituting free occurrences
of variable $x$ with $V$.

A call-by-value computation consists of successively applying the
$\beta_v$ reduction rule to a subterm. Positions of $\beta_v$ redexes
are restricted by an evaluation context which is defined as follows:
$$H  ::=  \Box \alt H M \alt V H $$
where $\Box$ represents a ``hole''. If $H$ is an evaluation context,
then $H[M]$ denotes the term that results from placing $M$ in the hole
of $H$. 
The evaluation of a program is then defined by a stepping relation,
denoted by $\mapsto$, given as follows: 
$$\displaystyle{\frac{M \rightarrow M'}{H[M] \mapsto H[M']}}$$


\subsection{Representation in \splname}
\label{subsec_slcbv}


\begin{figure}
\begin{textcode}
\begin{verbatim}
SIGNATURE:
type M = Var of string | Lam of string*M | App of M*M;;
startfrom M;;

SPECIFICATION:
#open "namesupply";;
let rec subst (t1,x,t2) = 
  match t1 with 
    Var s -> if s = x then t2 else t1
  | Lam(s,t1') -> if s = x then t1
      else let s' = freshname() in 
           Lam(s', subst(subst(t1',s,Var s'),x,t2))
  | App(t11,t12) -> App(subst(t11,x,t2),subst(t12,x,t2));;

dynamic V = Lam _;; 

axiom betav: App(Lam(x,t1), (t2:V)) ==> subst(t1, x, t2);;

context H =  BOX | App(H,_) |  App(V,H);;

inference eval: 
t1 ==> t2
-------------------
(h:H) t1 |==> h t2 ;;
\end{verbatim}
\end{textcode}
\caption{An \splname\ Specification of a Simple CBV Language}
\label{fig_slcbv}
\end{figure}

The specification of the
call-by-value $\lambda$-calculus in {\splname} is given in
Fig.~\ref{fig_slcbv}.
The \verb.SIGNATURE. part describes the abstract syntax of the
language using {\splsrchost} type definitions. In general, the
{\splname} type definitions 
may be polymorphic but are restricted to first-order type expressions
(no function types). To account for cases in which the description
needs more than one type, 
the type of programs in the object language is
explicitly given by the \verb|startfrom| phrase.

The \verb.SPECIFICATION. part describes the semantics of the language.
A dynamic definition defines a subset of a type with a
semantic significance. The axioms are
conditional rewriting rules. The optional conditions are {\splsrchost}
expressions following the keyword \verb|when|. 
The meta-language also has a primitive notion of contexts
with \verb.BOX. as the empty context. Each
inference rule has one premise clause and one conclusion clause, also with an
optional condition expression. 
Axioms and inference rules use a richer notion of pattern-matching
than the one 
used in most functional languages: they include dynamic constraints
like in \verb.t2:V.,  context constraints and context fillings like
in \verb.h:H. and \verb.h t2.. Meta-operations of the semantics like
substitution are written directly in {\splsrchost} as auxiliary
definitions. 

\subsection{Generating interpreters}

The \splname\ system is \emph{very} domain-specific,
targeting exactly the kind of semantic specifications based on
syntactic theories. 
In
addition, it performs basic checks to ensure the specifications are
well-formed. Other than the basic syntactic checks, the \splname\
system has a (meta-)type system that ensures that contexts are used
appropriately, \eg, every context has one hole, and contexts are
filled with expressions of the appropriate types, and both sides
of each axiom have the same type.  After performing these basic
checks, the \splname\ system compiles the specification into a 
non-deterministic automaton, which is then transliterated into
{\spltarget} code; 
the code uses success continuations for encoding the sequencing of
states; and exceptions with handlers for encoding the
non-deterministic selection of a state. 

Feeding the code in Fig.~\ref{fig_slcbv} to the \splname\ system
produces an interpreter. This interpreter can then be invoked on terms
of the language to evaluate them by repeatedly decomposing them into
evaluation contexts and redexes, and contracting the redexes, until
an answer is reached. For example, if the input file contains:

\begin{textcode}
\begin{verbatim}
App(Lam("y",Var "y"),App(Lam("x",Var "x"),Lam("z",Var "z")));;
\end{verbatim}
\end{textcode}

\noindent the generated interpreter produces:

\begin{textcode}
\begin{verbatim}
App(Lam("y",Var "y"),App(Lam ("x",Var "x"),Lam("z",Var "z")))
 ==>    by betav,eval
App(Lam("y",Var "y"),Lam("z",Var "z"))
 ==>    by betav,eval
Lam("z",Var "z")
\end{verbatim}
\end{textcode}

Interpreters generated by the {\splname} system preserve the semantics
of object-languages in the sense that if a specification is
non-deterministic, the generated interpreter evaluates input programs
non-deterministically. The current version does not employ
backtracking in evaluation.

\section{Compiling {\splname} specifications}
\label{sec_compile}

The compilation of an {\splname} specification includes the usual phases such
as lexing, 
parsing, static checking, and code generation. 
For an object-language specified by an {\splname} program, a parser and
a pretty-printer of the 
object-language are generated from the signature part, and a reduction machine
based on pattern-matching automata is generated from the semantic rules. 
These parts work together as an interpreter for the
object-language with the support of the {\splname} runtime libraries. 

Next, we introduce some issues in the {\splname} compilation such as
typing contexts, building 
automata, and transforming automata into {\spltarget} code.

\subsection{Typing contexts}

The type system for {\splname} extends the type system for
{\splsrchost}. 
The extensions deal with dynamic definitions and
contexts. Here,   
we only present the idea of typing contexts in a simply-typed
framework.
Typing dynamic definitions is similar.

First, we give definitions of meta-expressions, context expressions,
and their types.
\[
\begin{array}{l@{\quad}rll}
\textit{Expressions}  & E & ::= & c_0 \alt c_1 E \alt (E,E) \alt
 x \alt \lambda x. E \alt E~E \alt N[E]\\
\textit{Context Expressions} & H & ::= & \Box \alt N \alt c_1 H \alt
 (H,E) \alt (E,H) \alt N[H] 
\\
\textit{Context Definitions} & L & ::= & N ~\ttbox{=}~ H ~\ttbox{|}~
 \cdots ~\ttbox{|}~ H
\\
\textit{Types} & T & ::= & a \alt T * T \alt T \arrow T 
\\
\textit{Context Types} & U & ::= & T \circarrow T
\end{array}
\]

We write $x$ for variables, $c_0$ for nullary constructors, $c_1$
for constructors of arity one, $a$
for constant types, and $N$ for context names. 
Note that the symbol ``$\ttbox{=}$'' and the symbol ``$\ttbox{|}$'' in
a context definition are symbols of the {\splname} language.
The expressions include context filling, and tuple 
expressions are represented as nested pairs.
Context expressions are distinguished from expressions, for they
always contains one hole. 
The type of a context expression has the form 
$t_1 \circarrow t_2$, where $t_1$ is the type for the hole and $t_2$
is the type of the whole expression if the hole is filled. 
For a context definition 
$N~\ttbox{=}~H_1~\ttbox{|}~\cdots~\ttbox{|}~H_n$,
each $H_i$ should have the same type 
as the context type of $N$.

The typing rules are given in Table~\ref{fig_typingcontexts}. $\Gamma$
is a basis for type checking, which contains type assignments for
constructors, variables, and context names. It has properties such as
\concept{weakening} where it may have unused assignments,
\concept{strengthening} where unused assignments can be removed,
\concept{permutation} where the order of assignments is irrelevant,
and \concept{contraction} where assignments can be used more than
once. The first part in the table is the set of rules for
expressions. Most  
are standard except the rule for context filling  which
is  similar to 
function application. The second part is the set of rules for context
expressions. 
These rules express the ``lifting'' of the context type constructor
$\circarrow$ whenever possible, so that context expressions preserve
context types.  
The rule for filling contexts with context expressions is similar to
function composition. If a context $N$ has type $\tau_1 \circarrow
\tau_2$ and a context expression $h$ has type $\tau_0 \circarrow \tau_1$,
then the context expression $N[h]$ has type $\tau_0 \circarrow
\tau_2$.

\begin{table}
\label{typing rules}
\begin{center}
\begin{tabular}{|c|}
\hline
\\
$
\begin{array}{c@{\quad}c}

\infer{}{\tjudge{\Gamma, c_0:\tau}{c_0}{\tau}} 
&
\infer{}{\tjudge{\Gamma, c_1:\tau_1 \arrow \tau_2}{c_1}{\tau_1 \arrow
\tau_2}}
\\
\\
\infer{}{\tjudge{\Gamma, x:\tau}{x}{\tau}}
&
\infer{}{\tjudge{\Gamma, N:\tau_1 \circarrow \tau_2}{N}{\tau_1
\circarrow \tau_2}}

\\
\\

\infer{
        \begin{array}{c*{2}{@{\hspace{0.2in}}c}}
                \tjudge{\Gamma}{c_1}{\tau_1 \to \tau_2}  &
                \tjudge{\Gamma}{e}{\tau_1}
        \end{array}
        }
        {\tjudge{\Gamma}{c_1~e}{\tau_2}}
&
\infer{
        \begin{array}{c*{2}{@{\hspace{0.2in}}c}}
                \tjudge{\Gamma}{e_1}{\tau_1} & 
                \tjudge{\Gamma}{e_2}{\tau_2}
        \end{array}
        }
        {\tjudge{\Gamma}{(e_1,e_2)}{\tau_1 * \tau_2}}

\\
\\

\infer{\tjudge{\Gamma, x:\tau_1}{e}{\tau_2}}
        {\tjudge{\Gamma}{\lambda x. e}{\tau_1 \to \tau_2}}
&
\infer{
        \begin{array}{c*{2}{@{\hspace{0.2in}}c}}
                \tjudge{\Gamma}{e_1}{\tau_1 \to \tau_2} & 
                \tjudge{\Gamma}{e_2}{\tau_1}
        \end{array}
        }
        {\tjudge{\Gamma}{e_1~e_2}{\tau_2}}

\\
\\
\infer{
        \begin{array}{c*{2}{@{\hspace{0.2in}}c}}
                \tjudge{\Gamma}{e_1}{\tau_0 \circarrow \tau_1} &
                \tjudge{\Gamma}{e_2}{\tau_0}
        \end{array}}
        {\tjudge{\Gamma}{e_1[e_2]}{\tau_1}}
\\
\\
\hline
\\
\infer{}{\tjudge{\Gamma}{\Box}{\tau \circarrow \tau}} 
&

\infer{
        \begin{array}{c*{2}{@{\hspace{0.2in}}c}}
                \tjudge{\Gamma}{c_1}{\tau_1 \to \tau_2} &
                \tjudge{\Gamma}{h}{\tau_0 \circarrow \tau_1}
        \end{array}}
        {\tjudge{\Gamma}{c_1~h}{\tau_0 \circarrow \tau_2}}
\\ 
\\
\infer{
        \begin{array}{c*{2}{@{\hspace{0.2in}}c}}
                \tjudge{\Gamma}{h}{\tau_0 \circarrow \tau_1} &
                \tjudge{\Gamma}{e}{\tau_2}
        \end{array}}
        {\tjudge{\Gamma}{(h,e)}{\tau_0 \circarrow \tau_1*\tau_2}}
&
\infer{
        \begin{array}{c*{2}{@{\hspace{0.2in}}c}}
                \tjudge{\Gamma}{e}{\tau_1} &
                \tjudge{\Gamma}{h}{\tau_0 \circarrow \tau_2}
        \end{array}}
        {\tjudge{\Gamma}{(e,h)}{\tau_0 \circarrow \tau_1*\tau_2}}
\\
\\
\infer{
        \begin{array}{c*{2}{@{\hspace{0.2in}}c}}
                \tjudge{\Gamma}{N}{\tau_1 \circarrow \tau_2} &
                \tjudge{\Gamma}{h}{\tau_0 \circarrow \tau_1}
        \end{array}}
        {\tjudge{\Gamma}{N[h]}{\tau_0 \circarrow \tau_2}}
\end{array}
$
\\
\\
\hline
\end{tabular}
\end{center}
\caption{The type checking rules for {\splname} expressions and
context expressions}
\label{fig_typingcontexts}
\end{table}

\subsection{Building pattern-matching automata}

Pattern-matching is the crucial part in compiling an {\splname}
specification. 
A naive way is to check a list of 
patterns one by one. 
The obvious drawback of this approach is
inefficiency. Tree-like automata~\cite{treeautomata} address the
efficiency issue. The 
matching proceeds by making branches of different constructors and
ascribing the list of patterns to those branches. This approach has the
disadvantage of space explosion. 
The combination of tree automata with failures is the basis for
the current implementations of most common functional
languages~\cite{MarangetPat94,Leroy-ZINC}. 
The pattern-matching of the {\splname} system follows
this approach, but 
the support of semantic notions and the
non-determinism of rewriting require extensions to the existing
algorithms. 


\begin{enumerate}
\item{{\splname} Patterns}


The {\splname} patterns include common patterns such as wildcard patterns, 
variable patterns, alternative patterns,
type constraint patterns, alias patterns, and tuple patterns. 
The {\splname} patterns are also enriched with dynamic constraint
patterns and 
context filling patterns. 
The dynamic constraint $(p:dynamic\_name)$ and
 context filling $(p:context\_name)~p_2$
require $p$ to be a wildcard pattern or a variable pattern.
The {\splname} patterns are formally defined as follows:

\[
\begin{array}{l@{\quad}rll}
\textit{Patterns}  & P & ::= & \_ \alt x \alt c_0 \alt c_1 P \alt 
P \ttbox{|} P \alt  P~\ttbox{as}~x \\
& & &  (P:\ttbox{type}~\tau) \alt (P,\ldots,P)  \alt \\
                 &   &     & (P:dynamic\_name) \alt \\
                & & & (P:context\_name) ~ P \\
\textit{Axioms} & A & ::= & P~\ttbox{when}~E~\ttbox{==>}~E \\
\textit{Inference Rules} & I & ::= &
\infer{E~\ttbox{==>}~Q}{P~\ttbox{when}~E ~\ttbox{|==>}~ E} \\ 
\textit{Dynamic Definitions} & D  & ::= &  P \\
\textit{Context Definitions} & H(x) & ::= & P \\
\end{array}
\]

\noindent
where $E$ denotes {\splsrchost} expressions, and $Q$ denotes
restricted {\splname} patterns which do not contain dynamic
constraints and context fillings. 

The {\splname} patterns are used in the left-hand sides  of
axioms and of conclusion clauses of inference
rules.  
Dynamic definitions can be considered as definitions of {\splname}
patterns, and context definitions are parametric patterns where the
parameters denote the holes. Both definitions can be recursively
defined. 

The dynamic definition, context
definition, and rules in Fig.~\ref{fig_slcbv} can be represented
in terms of {\splname} 
patterns as follows: 
\[
\begin{array}{rcl}
V & = & \ttbox{Lam}~\_ 
\\
\beta_v & : & \ttbox{App}(\ttbox{Lam}(x,M), (v:V)) \ttbox{==>} M[x:=v]
\\
H(x) & = & x \mid \ttbox{App}((h_1:H)(x),\_) \mid
           \ttbox{App}((v:V),(h_2:H)(x)) 
\\
eval & : & \infer{t_1 ~\ttbox{==>}~ t_2}{(h:H)~t_1 ~\ttbox{|==>}~ h~t_2}

\end{array}
\]

where variables are introduced for dynamic constraint and
context filling patterns.

\item{Automata}

An automaton consists of states. Some states are
final. 
Matching a term consists of traversing the states until reaching a
final one. 
States are inductively defined as follows: 

\[
\begin{array}{l@{\qquad}rll}
\textit{States}  & S & ::= &  \bfbox{branch}~(t, (test,S),\ldots,(test,S))
     \alt \bfbox{accept}~ E \alt \\
& & & \chcstates{S}{S} \alt \bfbox{fail} \alt 
     \ttbox{let}~vars = f~E ~\ttbox{in}~S \alt \\
& & & \ttbox{if}~E~\ttbox{then}~S~\ttbox{else}~S \alt
     \ttbox{let}~vars = vars~\ttbox{in}~S 
\end{array}
\]

where $vars$ denotes a variable or a tuple of variables.

The first four states are standard. 
$\bfbox{branch}~(t,(test_1,S_1),\ldots,(test_n,S_n))$ is a branch-test
state, where $t$ is the term under test, and
each $test_i$  has the form $c_0$, $c_1~ x$, or
``\ttbox{\_}'' for otherwise. All tests are mutually disjoint.
$\bfbox{accept}~e$ is a final state, where $e$ is a {\spltarget}
expression representing the 
action after acceptance. A choice state
$\chcstates{S_1}{S_n}$ has alternatives $S_1,\ldots,S_n$.
When pattern-matching traverses this state, it non-deterministically chooses to enter
an alternative. If a final state is reached, then traversing the
choice state is successful. Otherwise it 
backtracks to other alternatives. 
This semantics differs from the usual choice state
whose alternatives are ordered (lexically).  
$\bfbox{fail}$ is a failure state.
One new form, reference state $\ttbox{let}~vars = f~e
~\ttbox{in}~S_1$, is added to support dynamic values and contexts. It
calls a function matching the parameter as the corresponding dynamic value,
context, or redex . If it succeeds, it continues in state $S_1$,
failures in $S_1$ may cause backtracking to other possibilities in the
function call $f~e$. 


States are also extended with conditional expressions and \ttbox{let}
variable bindings. The reason for the former
extension is that the  
semantic rules are conditional. The latter extension is helpful for
code generation. 
States are annotated with terms to be matched, but we made them
implicit in our presentation. 

\item{Structures for pattern-matching}

The {\splname} compiler collects the inference rules and axioms with the
same type together. 
The patterns of the rules form a vector which can be considered as
a one-column matrix. 
Each rule contributes a row in the matrix.
We introduce
parameters bound to the terms matching the patterns, and we
keep track of 
variable bindings in pattern-matching. There is also
a state for each rule, indicating what to do when the
patterns of the rule have been matched. 
The whole pattern-matching structure is represented as
follows:    

\[
\left(
\begin{array}{c@{\cdots}cc}
t_1 & t_n \\
p_{11}  & p_{1n}  & s_1 \\
\vdots  & \vdots  & \vdots \\
p_{m1}  & p_{mn}  & s_m
\end{array}
\right)
\]

The compilation of pattern-matching can be regarded as a function,
${\cal C}$, which maps such a structure to a state.


The initialization of the pattern-matching  sets the states in
the structure. 
\begin{itemize}
\item
For an axiom $p_1 ~\ttbox{when}~ e_c ~\ttbox{==>}~ e_r$, the
 corresponding state
 is: \\
$$
\ttbox{if}~e_c~\ttbox{then}~\bfbox{accept}~e_r~\ttbox{else}~\bfbox{fail}$$
\item 
For an inference rule $\infer{e_1 ~\ttbox{==>}~ p_1}{p_2 ~\ttbox{when}~
e_c ~\ttbox{==>}~ e_2}$, the corresponding state  is: 
$$
\ttbox{if}~e_c~\ttbox{then}~\ttbox{let}~p_1 = \ttbox{rewrite1}~
e_1~\ttbox{in}~\bfbox{accept}~e_2~\ttbox{else}~\bfbox{fail}$$
where $\ttbox{rewrite1}$ is one-step rewriting function in the
generated code.  
\end{itemize}

\item{Pattern-matching algorithm}

The pattern-matching algorithm is a divide-and-conquer algorithm. It
selects one column 
of the pattern matrix to work on according to certain
criteria. Without loss of 
generality, we assume that the algorithm always chooses the first
column. The algorithm repeats the following steps until the pattern
matrix is empty:

\begin{enumerate}

\item Preprocessing 

This step canonicalizes the patterns in the first column. It
removes the type constraints since the type information is not useful
at the current stage. It binds variables in alias patterns to the
corresponding parameters. It turns each alternative pattern
into several rows having one alternative each. 
Formally, the preprocessing repeats
the following simplifications. 

\[
\begin{array}{lcl}
{\cal C}\left(
\begin{array}{c@{\cdots}c}
t_1 &  \\
\vdots  & \vdots \\
(p_{i1}:\ttbox{type}~\tau)  & s_i \\
\vdots  & \vdots 
\end{array}
\right)
&
\longrightarrow
&
{\cal C}\left(
\begin{array}{c@{\cdots}c}
t_1 \\
\vdots  & \vdots \\
p_{i1}  & s_i \\
\vdots  & \vdots 
\end{array}
\right)
\\
{\cal C}\left(
\begin{array}{c@{\cdots}c}
t_1  \\
\vdots  & \vdots \\
(p_{i1}~\ttbox{as}~x)  & s_i \\
\vdots  & \vdots 
\end{array}
\right)
&
\longrightarrow
&
{\cal C}\left(
\begin{array}{c@{\cdots}c}
t_1 \\
\vdots  & \vdots \\
p_{i1}  & \letstate{x=t_1}{s_i} \\
\vdots  & \vdots 
\end{array}
\right)
\\
{\cal C}\left(
\begin{array}{c@{\cdots}c}
t_1 \\
\vdots  & \vdots \\
p_{i1a} \ttbox{|} p_{i1b}  & s_i \\
\vdots  & \vdots 
\end{array}
\right)
&
\longrightarrow
&
{\cal C}\left(
\begin{array}{c@{\cdots}c}
t_1 \\
\vdots  & \vdots \\
p_{i1a}  & s_i \\
p_{i1b}  & s_i \\
\vdots  & \vdots
\end{array}
\right)
\end{array}
\]

\item Splitting the matrix

The algorithm splits the matrix horizontally, so that in each submatrix,
all first-column patterns are in one of the following groups:
\begin{itemize}
\item \concept{variable group}: wildcard patterns or variable patterns,
\item \concept{tuple group}: tuple patterns, 
\item \concept{constructor group}: constructor patterns,
\item \concept{dynamic constraint group}: dynamic constraint patterns
on the same dynamic definition,
\item \concept{context filling group}: context filling patterns on the
same context definition. 
\end{itemize}
The result of the splitting is a choice state. 

\[
\begin{array}{rcl}
{\cal C}\left(
\begin{array}{c@{\cdots}c}
t_1 \\
p_{11} & s_1 \\
\vdots & \vdots \\
p_{m1} & s_m
\end{array}
\right) 
&
\longrightarrow
&

\chcstates
{{\cal C}\left(
\begin{array}{c@{\cdots}c}
t_1 \\
p_{11} & s_1 \\
\vdots & \vdots \\
p_{{k_1}1} & s_{k_1}
\end{array}
\right)~}
{~{\cal C}\left(
\begin{array}{c@{\cdots}c}
t_1 \\
p_{{(k_l+1)}1} & s_{k_l+1} \\
\vdots & \vdots \\
p_{m1} & s_m
\end{array}
\right)}

\end{array}
\]

\item Analyzing different cases  

For each submatrix, the compilation function ${\cal C}$ is inductively
defined as follows:

\begin{enumerate}

\item Base case:

When the pattern matrix is empty, pattern-matching is vacuously
successful. The 
function ${\cal C}$ creates a choice state.

\[
\begin{array}{lcl}
{\cal C}\left(
\begin{array}{c}
s_1 \\
\vdots \\
s_m \\
\end{array}
\right)
&
\longrightarrow
&
\chcstates{s_1}{s_m}
\end{array}
\]

\item Variable group:

Wildcard  and variable patterns always match successfully.
The function ${\cal C}$ removes the column of patterns.
For variable patterns, bindings are added for further access to
the variables.

\[
\begin{array}{lll}
{\cal C}\left(
\begin{array}{c@{\cdots}c}
t_1 \\
\vdots & \vdots \\
\_ & s_i \\
\vdots & \vdots \\
x & s_j \\
\vdots & \vdots 
\end{array}
\right) 
&
\longrightarrow
&

{\cal C}\left(
\begin{array}{c@{\cdots}c}
t_2 \\
\vdots & \vdots \\
p_{i2} & s_i \\
\vdots & \vdots \\
p_{j2} & \letstate{x=t_1}{s_j} \\
\vdots & \vdots 
\end{array}
\right)

\end{array}
\]

\item Tuple group:

The function ${\cal C}$ treats each component of the tuple as
an individual pattern. It  replaces the first 
column of patterns with columns of the component patterns, and
introduce parameters for the components. 

\[
\begin{array}{rcl}
{\cal C}\left(
\begin{array}{c@{\cdots}c}
t_1 \\
(p_{111},\cdots,p_{11k}) & s_1 \\
\vdots & \vdots \\
(p_{m11},\cdots,p_{m1k}) & s_m
\end{array}
\right) 
&
\longrightarrow
&
\begin{array}{l}
\letstate{(t_{11},\cdots,t_{1k})=t_1}{} \\
\quad
{
{\cal C}\left(
\begin{array}{c@{\cdots}c@{\cdots}c}
t_{11}& t_{1k}  \\
p_{111} & p_{11k} & s_1 \\
\vdots & \vdots & \vdots \\
p_{m11} & p_{m1k} & s_m
\end{array}
\right)
}
\end{array}
\end{array}
\]

\item Constructor group:

The function ${\cal C}$ collects the rows which have the same 
constructor in the first column into a group of new
pattern-matching structures, and it creates a
branch-test state. The tests of the state are distinguished by having
different constructors as roots.
The corresponding
actions for the tests are the results of compiling the new
structures. 
In the new structures, the first column is removed for the
constructors of zero arity, or it is replaced by the column of
argument patterns for the constructors of non-zero arity.


\[
\begin{array}{rcl}
{\cal C}\left(
\begin{array}{c@{\cdots}c}
t_1 \\
\vdots & \vdots \\
c_0 & s_i \\
\vdots & \vdots \\
c_1~p'_{j1} & s_j \\
\vdots & \vdots 
\end{array}
\right) 
& 
\longrightarrow

&

\bfbox{branch}
\left(
t_1,
\begin{array}{ll}
(c_0, &
{\cal C}\left(
\begin{array}{c@{\cdots}c}
t_2  \\
p_{{i_1}2} & s_{i_1} \\
\vdots & \vdots \\
p_{{i_2}2} & s_{i_2}
\end{array}
\right))
\\
\quad\vdots & \quad\quad \vdots
\\
(c_1~t'_1, & 
\begin{array}{l}
\letstate{c_1~t'_1=t_1}{} \\
\quad
{
{\cal C}\left(
\begin{array}{c@{\cdots}c}
t'_1 \\
p'_{{j_1}1} & s_{j_1} \\
\vdots & \vdots \\
p'_{{j_2}1} & s_{j_2}
\end{array}
\right)
}
\end{array}
)
\end{array}
\right)
\end{array}
\]

\item Dynamic constraint group:

The function ${\cal C}$ creates a reference state.  
The reference will initiate pattern-matching for the
dynamic definition $D$ with the value of $t_1$. 
Its result is bound to a new parameter.
The state in the $\ttbox{let}$ body is the result of 
compiling a structure consisting of the patterns without
the constraint and the rest of the patterns.

\[
\begin{array}{rcl}
{\cal C}\left(
\begin{array}{c@{\cdots}c}
t_1 \\
(p'_{11}: D) & s_1 \\
\vdots & \vdots \\
(p'_{m1}: D) & s_m
\end{array}
\right) 
&
\longrightarrow
&
\begin{array}{l}
\ttbox{let}~ t'_1 = \ttbox{match\_}D ~t_1 ~\ttbox{in} \\
\quad
{\cal C}\left(
\begin{array}{c@{\cdots}c}
t'_1 \\
p'_{11} & s_1 \\
\vdots  & \vdots \\
p'_{m1} & s_m
\end{array}
\right)
\end{array}

\end{array}
\]

Pattern-matching for dynamic definitions will be presented later.

\item Context filling group

Similar to the dynamic constraint group, the function 
${\cal C}$ creates a reference state.  The reference will
initiate pattern-matching for the context definition $H$ with
the value of $t_1$. 
Its result is bound to a pair of parameters which represent the context and the corresponding hole occurring in $t_1$.
The state of the $\ttbox{let}$ body is the result of 
compiling a structure consisting of the context patterns, 
the hole patterns,
as well as the rest of the patterns. 

\[
\begin{array}{rcl}
{\cal C}\left(
\begin{array}{c@{\cdots}c}
t_1 \\
(p'_{11}: H) p''_{11} & s_1 \\
\vdots & \vdots \\
(p'_{m1}: H) p''_{m1} & s_m
\end{array}
\right) 
&
\longrightarrow
&
\begin{array}{l}
\ttbox{let}~ (t'_1,t''_1) = \ttbox{match\_}H ~t_1 ~\ttbox{in} \\
\quad
~{\cal C}\left(
\begin{array}{cc@{\cdots}c}
t'_1 & t''_1 \\
p'_{11} & p''_{11} & s_1 \\
\vdots & \vdots & \vdots \\
p'_{m1} & p''_{m1} & s_m
\end{array}
\right)
\end{array}

\end{array}
\]

Pattern matching for context definitions will be presented next.

\end{enumerate}

\end{enumerate}

\item Matching dynamic definitions and context definitions

Pattern-matching for dynamic definitions and context definitions uses
the same form of 
structures and uses the same algorithm. Each definition corresponds to
a structure which has only one pattern, one parameter and one
state. The pattern comes from the definition. Assume $t$ is the term
to be matched, the state is initialized
as follows:  
\begin{itemize}
\item 
For a dynamic definition, the state is $\bfbox{accept}~t$.
\item 
For a context definition, the state is $\bfbox{accept}~(\lambda
x. body, x)$, where $x$ is the parameter for the context and the $body$
is a placeholder. When the base case in the algorithm is encountered, the
{\splname} compiler sets the contents of $body$ by reconstructing the
term $t$ with the variable $x$. Each $body$ may have different content
if the final state is copied. The reconstruction
retrieves the bindings along the path from the start state to the 
final state. 

In other words, matching a context definition returns a pair with a
constructing function and a term filling the hole. Applying the
function to the 
term results in the  term $t$.

\end{itemize}

The state starting pattern-matching for a definition can be referred
to by the name of the definition. For example, the dynamic definition
 and the context definition in
 Fig.~\ref{fig_slcbv} are associated with the following states:
\[
\begin{array}{rcl}
\ttbox{match\_}D ~t & = & 
\bfbox{branch}(t,(\ttbox{Lam}~t', \bfbox{accept}(t)))
\\
\\
\ttbox{match\_}H ~t & = & 
\begin{array}[t]{l}
\ttbox{let}~ x = t ~\ttbox{in}~ \bfbox{accept}(\lambda x.body_1,x) \quad \chcstatessep
\\
\bfbox{branch}(t,
\\
\quad
(\ttbox{App}~t', 
\begin{array}[t]{l}
\ttbox{let}~ (t_1,t_2) = t' ~\ttbox{in} 
\\
\quad 
\begin{array}{l} 
(\ttbox{let}~(t'_1,t''_1) = \ttbox{match\_}H ~t_1 ~\ttbox{in} 
\\
 \quad \ttbox{let}~ h_1 = t'_1 ~\ttbox{in} 
\\
 \quad\quad \ttbox{let}~ x = t''_1 ~\ttbox{in}~
\bfbox{accept}(\lambda x.body_2, x)) \quad \chcstatessep
\end{array}
\\
\quad
\begin{array}{l}
(\ttbox{let}~t'_1 = \ttbox{match\_}D ~t_1 ~\ttbox{in}~
 \ttbox{let}~ v = t'_1 ~\ttbox{in}~ 
\\
 \quad
\begin{array}{l}
\ttbox{let}~(t'_2,t''_2) = \ttbox{match\_}H ~t_2 ~\ttbox{in} 
\\
 \quad \ttbox{let}~ h_2 = t'_2 ~\ttbox{in}
\\
 \quad\quad \ttbox{let}~ x=t''_2 ~\ttbox{in}~ 
 \bfbox{accept}(\lambda x.body_3, x))))
\end{array}
\end{array}
\end{array}
\end{array}
\end{array}
\]

where $body_1$, $body_2$ and $body_3$ are as follows:
\[
\begin{array}{rcl}
body_1 & : & 
\ttbox{let $t$ = $x$ in $t$}
\\
body_2 & : & 
\ttbox{let $t''_1$ = $x$ in let $t'_1$ = $h_1$ in let $t_1$ =
$t'_1$ $t''_1$ in} \\
& & \quad \ttbox{let $t'$ = ($t_1$,$t_2$) in let $t$ = App $t'$ in
$t$}
\\
body_3 & : & 
\ttbox{let $t''_2$ = $x$ in let $t'_2$ = $h_2$ in let $t_2$ =
$t'_2$ $t''_2$ in} \\
& & \quad \ttbox{let $t'_1$ = $v$ in let $t_1$ = $t'_1$ in let $t'$ =
($t_1$,$t_2$) in} \\
& & \qquad \ttbox{let $t$ = App $t'$ in $t$}
\end{array}
\]

\end{enumerate}

\subsection{Transforming automata into {\spltarget} code}

The transformation of states in the
pattern-matching automata into {\spltarget} code is described in 
Fig. \ref{fig_codegen}.
Each state  corresponds to a function with implicit parameters
for the terms to be matched
and a continuation expressing the remaining pattern-matching work. 
The initial continuation for rewriting is the identity function.
State
functions may raise an exception when matching of the state fails. 
The branch-test state
corresponds to the {\tt match $\cdots$ with $\cdots$} construct in
{\splsrchost}. If the branch tests do not cover all
constructors of the same type, a default test associated with a
failure state is added. 
For a final state, the success continuation is consumed by applying it
to the expression. 
For a choice state, a random alternative is tried. Failure exceptions
may be caught and then other alternatives can be tried. 
The failure state just raises a failure
exception. 
For a reference state, it calls the matching function with the
continuation accepting the result, then it continues with the state in
the body of the $\ttbox{let}$.
The conditional expressions and variable bindings in states are
considered atomic 
with respect to continuation passing and exception handling. 
Transforming a conditional state is thus transforming both
branch states, and transforming  a $\ttbox{let}$ state 
is transforming the $\ttbox{in}$ part. 

\begin{figure}

\[
\begin{array}{l}
\transstate{\bfbox{branch } (t, (c_0, S_0), \ldots,(c_1~x, S_1))}{k} = \\
\qquad \ttbox{match } t ~\ttbox{with} \\
\qquad
\begin{array}{lcl}
c_0 & \ttbox{->} & \transstate{S_0}{k} \\
& \vdots \\
c_1~x & \ttbox{->} & \transstate{S_1}{k} \\
\ttbox{\_ } & \ttbox{->} & \ttbox{raise failure}
\end{array}
\\
\transstate{\bfbox{accept } e}{k} = k~(e) \\
\transstate{\chcstates{S_1}{S_m}}{k} = \\
\qquad \ttbox{try } \transstate{S_i}{k}
        \hspace{4cm} 0\leq i \leq m\\
\qquad \ttbox{with failure -> } \\
\qquad\qquad \transstate{\chcstates{S_1}{S_{1-i}}~\chcstatessep~\chcstates{S_{i+1}}{S_m}}{k} \\
\transstate{\bfbox{fail}}{k} = \ttbox{raise failure} \\
\transstate{\ttbox{let}~vars = f~e ~\ttbox{in}~S}{k}
        = \\
\qquad f~e ~(\lambda t. ~\ttbox{let}~vars = t ~\ttbox{in}~ \transstate{S}{k})
\\
\transstate{\ttbox{if}~e_c~\ttbox{then}~S_1~\ttbox{else}~S_2}{k}
        = \\
\qquad \ttbox{if } e_c ~\ttbox{then } \transstate{S_1}{k}
        \ttbox{else } \transstate{S_2}{k} \\
\transstate{\ttbox{let}~vars_1 = vars_2 ~\ttbox{in}~S}{k}
        = \\
\qquad \ttbox{let } vars_1 = vars_2 ~\ttbox{in}~ \transstate{S}{k}
\end{array}
\]

\caption{Generating {\spltarget} code}
\label{fig_codegen}
\end{figure}

\section{Summary and Discussion}
\label{sec_discussion}

The {\splname} system uses the first-order types of functional
languages for specifying abstract syntax.
We have not followed the approach of \emph{higher-order
abstract syntax (HOAS)}~\cite{hoas}. The HOAS 
representation would allow the
variables of the object language to be represented as 
meta-variables, and hence alleviates the need for explicitly reasoning
about variable renaming and substitution of variables. However,
higher-order abstract syntax interacts poorly with the inductive
reasoning techniques needed to reason about the properties of semantic
specifications~\cite{rechoas}.

The {\splname} system uses conditional
rewriting rules for specifying semantic rules. In other words, the
system employs rewriting semantics ({\aka} reduction semantics). Some
systems\cite{Hannan, Centaur88} express rules in
natural semantics\cite{NatSem}. 
Natural semantics is well adapted to describing static
behaviors such as typing.  For transformational behaviors, such as
dynamic semantics, rewriting semantics proves to be more 
modular\cite{WrightFelleisen94}, and
should therefore be more tractable when it comes to expressing
full-size programming languages.
Another advantage of rewriting semantics is that it allows one 
to observe intermediate states of reductions, so that
it is more suitable for non-terminating object-systems. 

ELAN\cite{ELANsystem} is a system also based on first-order rewriting
semantics. It 
has more general application areas than the {\splname} system. 
It supports many-to-one associative-commutative(AC) patterns and
provides an efficient algorithm\cite{MoreauK-PLILP+ALP98}.
Its strategy language gives flexible control over non-deterministic
reductions. The
Stratego\cite{VB98} system has more generic strategy
specifications. There are also  general-purpose tools that can be
used for manipulating formal semantics, such as Coq\cite{Coqsystem},
Isabelle\cite{Isabellesystem} and Twelf\cite{Twelfsystem}. Compared to
those systems, the novelty of the 
{\splname} system is that it directly supports the specification
of the semantic notions such as 
dynamic values and evaluation contexts, and it automatically generates
executable interpreters.  Associative-commutative patterns can be
represented by contexts, but  
the current {\splname} system does not optimize matching for efficiency. 
 
Fahndrich and Boyland~\cite{pattern-abstractions} investigate
abstract patterns which are even more general than contexts in the
{\splname} system.  They
build automata for patterns 
and check the properties of the patterns such as exhaustiveness and
non-overlapping, but they did not provide a computational view of
pattern-matching.

The current prototype of the {\splname} system constitutes a first
step towards  
designing a specification language for syntactic theories and 
implementing a system generating interpreters from specifications. It
also provides an extension to pattern-matching techniques. 
A number of interesting examples have been tested in the prototype system.
Follow-up work is underway to 
 allow more expressive specifications,
to optimize the compilation of the {\splname} language, and to
automatically prove properties of syntactic theories.

\bigskip
\noindent
{\bf Acknowledgment:} 
The first author thanks the Cristal group at INRIA Rocquencourt for
funding his visit from July,1998 to September, 1998. We thank
Xavier Leroy, Didier Remy, Amr Sabry, and Miley Semmelroth for
stimulating discussions about this work. 
\bigskip

\bibliographystyle{alpha}
\bibliography{general,reference}

\end{document}